\documentclass[runningheads]{llncs}
\usepackage[misc,geometry]{ifsym}
\usepackage{graphicx}
\usepackage{amsmath}
\usepackage{amssymb}
\usepackage{booktabs}
\usepackage[ruled,vlined]{algorithm2e}
\usepackage{epsfig}
\usepackage{physics}
\usepackage[inline]{enumitem}
\usepackage{comment}
\usepackage{booktabs,framed}
\usepackage{multirow}
\usepackage{xcolor,soul}
\usepackage{latexsym,url}
\usepackage{stackengine} 
\usepackage{wrapfig}

\usepackage[pagebackref,breaklinks]{hyperref}
\usepackage[capitalize]{cleveref}
\crefname{section}{Sec.}{Secs.}
\Crefname{section}{Section}{Sections}
\Crefname{table}{Table}{Tables}
\crefname{table}{Tab.}{Tabs.}

\newcommand{\myfirstpara}[1]{\par \noindent \textbf{#1:}}
\newcommand{\mypara}[1]{ \myfirstpara{#1}}

\newcommand{\beginsupplement}{%
        \setcounter{table}{0}
        \renewcommand{\thetable}{S\arabic{table}}%
        \setcounter{figure}{0}
        \renewcommand{\thefigure}{S\arabic{figure}}%
     }
     
\begin{document}
\title{Unsupervised Contrastive Learning of \\ Image Representations from Ultrasound Videos \\ with Hard Negative Mining}
\titlerunning{Contrastive Learning from USG Videos}
%
%
\author{
    Soumen Basu \inst{1} \Letter ~ \orcidID{0000-0002-3915-7545} 
    \and
    Somanshu Singla\inst{1} 
    \and
    Mayank Gupta\inst{1} 
    \and
    Pratyaksha Rana\inst{2}
    \and
    Pankaj Gupta\inst{2}
    \and
    Chetan Arora \inst{1}
}

%
\authorrunning{S. Basu et al.}
%
\institute{
Indian Institute of Technology, Delhi, India \\
\email{soumen.basu@cse.iitd.ac.in}
\and
Postgraduate Institute of Medical Education and Research, Chandigarh, India
}
\maketitle              
%

%
\begin{abstract}
	Rich temporal information and variations in viewpoints make video data an attractive choice for learning image representations using unsupervised contrastive learning (UCL) techniques. State-of-the-art (SOTA) contrastive learning techniques consider frames within a video as positives in the embedding space, whereas the frames from other videos are considered negatives. We observe that unlike multiple views of an object in natural scene videos, an Ultrasound (US) video captures different 2D slices of an organ. Hence, there is almost no similarity between the temporally distant frames of even the same US video. In this paper we propose to instead utilize such frames as hard negatives. We advocate mining both intra-video and cross-video negatives in a hardness-sensitive negative mining curriculum in a UCL framework to learn rich image representations. We deploy our framework to learn the representations of Gallbladder (GB) malignancy from US videos. We also construct the first large-scale US video dataset containing 64 videos and 15,800 frames for learning GB representations. We show that the standard ResNet50 backbone trained with our framework improves the accuracy of models pretrained with SOTA UCL techniques as well as supervised pretrained models on ImageNet for the GB malignancy detection task by 2--6\%. We further validate the generalizability of our method on a publicly available lung US image dataset of COVID-19 pathologies and show an improvement of 1.5\% compared to SOTA. Source code, dataset, and models are available at \url{https://gbc-iitd.github.io/usucl}.
	
\end{abstract}

\begin{figure}[t]
    \centering
    \includegraphics[width=\linewidth]{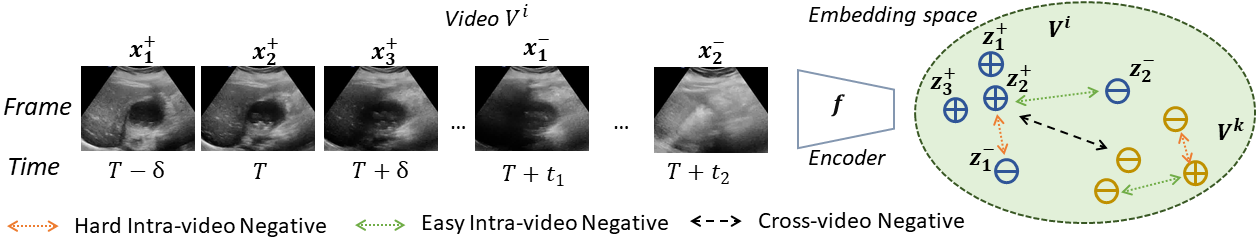}
    \caption{We motivate the use of intra-video negatives in contrastive learning. Based on the visibility of a pathology in the intra-video samples, negatives can be sampled. The frames in range $[T-\delta, T+\delta]$ in video $V^i$ has stones and malignant wall thickening visible for a small $\delta$. A slightly distant frame $T+t_1$ shows a GB, but the malignant wall thickening is not visible. This sample acts as a hard negative. The viewing plane further changes in frame $T+t_2$ and GB becomes invisible.}
    \label{fig:teaser}
\end{figure}

%
\section{Introduction}

Due to their remarkable performance, Deep Neural Networks (DNNs) have become defacto standard for a wide range of medical image analysis tasks in recent years \cite{ardila2019end,bejnordi2017diagnostic}. However, lack of annotated medical data due to the specialized nature of annotations, and the data privacy issues restrict the applicability of supervised learning of DNNs in medical imaging. Although pretraining on large natural image datasets yields a performance boost for downstream tasks on medical data \cite{alzubaidi2020transferlearning,cheng2017transfer}, the large domain gap between the natural and medical images remains a bottleneck. Recent works are increasingly going beyond the supervised setup and exploiting unsupervised techniques to compensate for the lack of annotated data \cite{simclr,moco}. Broadly, there are two prominent categories of representation learning techniques for leveraging unlabeled data. In the pretext task-based method, the DNNs are pretrained with some spatial tasks such as image rotation prediction \cite{komodakis2018unsupervised} or temporal tasks such as video clip order prediction \cite{xu2019self} to learn efficient image representation. On the other hand, contrastive methods \cite{simclr} try to distinguish between different views of image samples to learn robust representation without labels. Visual representations learned by contrastive learning have been shown to outperform the supervised pretraining on large annotated data in terms of accuracy on the downstream prediction tasks \cite{simclr,moco}.  

Video data contains rich variations in viewpoints and natural temporal information for objects making it suitable for going beyond image-level contrastive learning. Additionally, the abundance of unlabeled video data makes it an attractive choice for learning representations. Recent works are attempting to exploit the video data to learn robust image-level representations \cite{uscl,cyclecontrast}. We observe that current SOTA techniques for image representation learning from videos, such as USCL \cite{uscl} suggest that images coming from the same video are too close to be considered negatives (\emph{similarity conflict}). USCL advocates considering only cross-video samples as negatives to avoid the similarity conflict. While similarity conflict is prevalent for intra-video samples of natural video datasets like action recognition, where each video contains a distinct action, the US videos are inherently different. The frames of a US video contain both types of images where pathology is visible or absent. We argue that frames from the same video where the pathology is absent can be used as hard negatives for the positive samples with the visible pathology. \cref{fig:teaser} shows the positive and negative frames from the same US video. The temporal distance between the frames acts as a proxy to the hardness. Negative samples that are temporally closer to the positives contain higher similarities with the positives and are harder to differentiate in the embedding space. We propose an unsupervised contrastive learning framework to exploit both the intra-video and cross-video negatives for learning robust visual representations from US videos. We design a hardness-sensitive negative mining curriculum to lower the distance between the anchor and negatives gradually. Due to its unsupervised nature, our technique can be used for pretraining a backbone on any US video dataset for superior downstream performance.

We deploy our video contrastive learning framework to pretrain a neural network before supervised fine-tuning to identify GB malignancy in US images. Due to its non-ionizing radiation, low cost, and accessibility, US is a popular non-invasive diagnostic modality for patients with suspected GB pathology. Although there are prior works involving DNNs to detect GB afflictions such as polyp or stones \cite{gbPolyp,gbPolyp2,gbAutomatic}, there is limited prior work on using DNNs to detect GB malignancy in US images \cite{basu2022surpassing}. Unlike detecting stones or polyps, identifying GB malignancy from the routine US is challenging for radiologists \cite{gb-rads-paper,gupta2020imaging}. We observed that ImageNet pretrained classifiers perform even worse than radiologists. We used an in-house abdominal US video dataset to pretrain our model. Our contrastive learning-based pretrained model surpasses human radiologists and SOTA contrastive learning methods on GB malignancy classification.

We also validate our framework on a public lung US dataset, POCUS \cite{pocus}, containing COVID-19 and Pneumonia samples. Pretraining our model on public lung US videos, and finetuning for the downstream classification shows improvement over ImageNet pretraining and the current SOTA contrastive techniques.

\mypara{Contributions} The key contributions of this work are:
\begin{itemize}
\itemsep0em
	\item  We design an unsupervised contrastive learning technique for US videos to exploit both cross-video and intra-video negatives for rich representation learning. We further use a hardness-sensitive negative mining curriculum to boost the performance of our contrastive learning framework.
	\item We deploy our framework to solve a novel GB malignancy classification problem from US images. We also validate the efficacy of our technique on a publicly available lung US dataset for COVID detection.
	\item We are contributing the first US video dataset of 64 videos and 15800 frames containing both malignant and non-malignant GB towards the development of representation learning from medical US videos.
\end{itemize}

%
\begin{figure}[t]
    \centering
    \includegraphics[width=\textwidth]{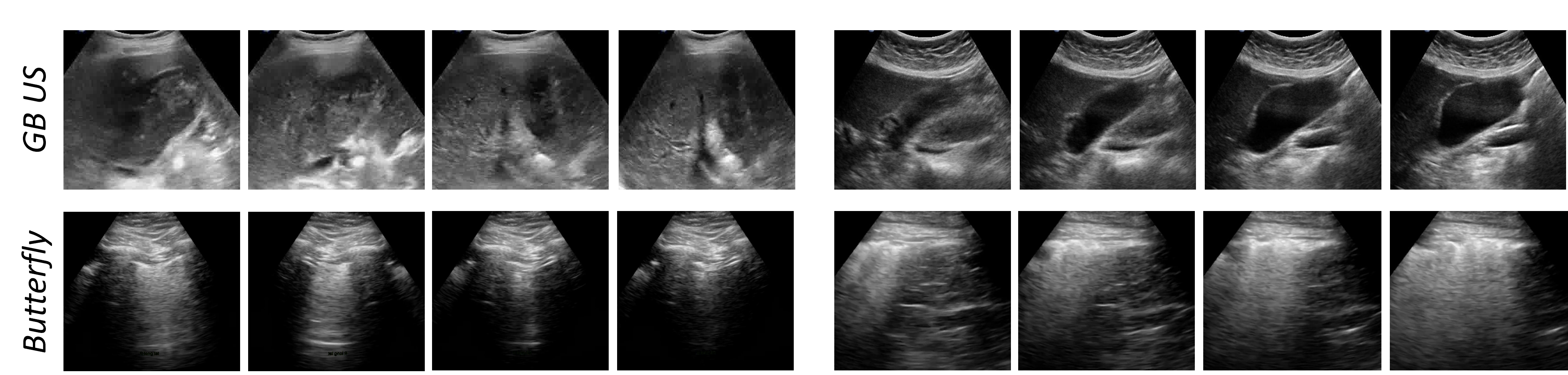}
    \caption{Sample video sequences from the GB US video and the Butterfly \cite{butterfly} datasets. Two sequences of size 4 is shown on the left and right for each dataset.}
    \label{fig:data_sample}
\end{figure}
\section{Datasets}
\subsection{In-house US Dataset for Gallbladder Cancer}
The transabdominal US video dataset is acquired at the Postgraduate Institute of Medical Education and Research (PGIMER), Chandigarh, India. The PGIMER ethics committee approved the study.

\mypara{Video Data}
Radiologists with 2-8 years of experience in abdominal sonography acquired the data. The US videos were obtained after at least 6 hours of fasting using a 1--5 MHz curved array transducer (C-1-5D, Logiq S8, GE Healthcare). The scanning intended to include the entire GB and the lesion or pathology. The frame rate was 29 fps. The length of the videos varied from 43 to 888 frames depending on the GB distension and size of the lesion. The dataset consists of 32 malignant and 32 non-malignant videos containing a total of 12,251 and 3,549 frames, respectively. Note that we do not use the video level labels in our setup. We cropped the video frames from the center to anonymize the patient information and annotations. The processed video frames were of size $360\!\times\!480$ pixels. \cref{fig:data_sample} shows some samples from the video data. We are releasing this large scale video dataset for the community.

\mypara{Image Data}
We use the publicly contributed GBCU dataset \cite{basu2022surpassing} consisting of 1255 US images from 218 patients. The dataset consists of 990 non-malignant (171 patients with normal and benign GB) and 265 malignant (47 patients) GB images. The images were labeled as normal, benign, or malignant, and such labels were biopsy-proven. We use this dataset for finetuning and report 10-fold cross-validation. 
Note that the patients recorded in the video dataset are not included in this image dataset to ensure generalization. 

\subsection{Public Lung US Dataset for COVID-19}

\myfirstpara{Video Data}
We use the public lung US video dataset, Butterfly \cite{butterfly}. Butterfly consists of 22 US videos containing 1533 images of size $658\!\times\!758$ pixels of the lung. The dataset was collected using a Resona 7T machine.

\mypara{Image Data}
We use the publicly available POCUS \cite{pocus} dataset consisting of a total of 2116 lung US images, of which 655, 349, and 1112 images are of COVID-19, bacterial pneumonia, and healthy control, respectively.

%
%
\section{Our Method}

\begin{figure}[t]
    \centering
    \includegraphics[width=0.9\linewidth]{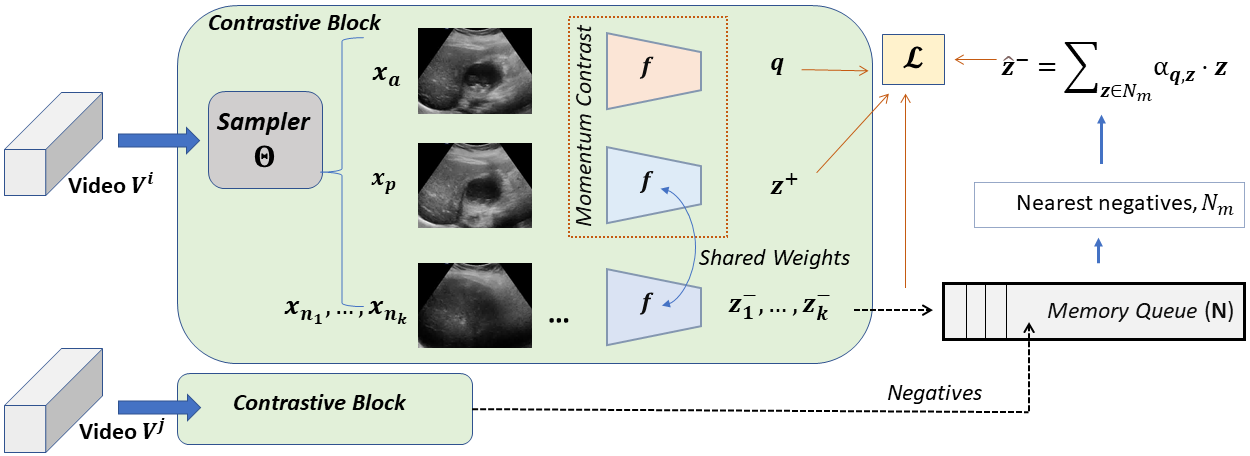}
    \caption{Overview of the proposed contrastive loss, $\mathcal{L}$. An anchor $\vb{q}$ and another temporally close sample $\vb{z}^+$ from the same video are used as positive pairs. Given the set of cross-video negatives $N$, and the intra-video negatives $\vb{z}_j^-$, we compute $\vb{\hat{z}}^-$ from the $m$ most similar cross-video negatives to the anchor. The intra-video samples $\vb{z}_j^-$ and $\vb{\hat{z}}^-$ are considered as negatives to the anchor.}
    \label{fig:my_label}
\end{figure}

\myfirstpara{Contrastive Learning Setup}
Suppose, $\vb{V}^i = \big\{ \vb{x}_j\big\}_{j=1}^{M^i}$ is the $i$-th video in the video dataset consisting of $M^i$ frames where $\vb{x}_j$ is the $j$-th individual frame. We sample an anchor image, $\vb{x}_a$, a positive image $\vb{x}_p$, and $k$ negative frames $\vb{x}_{n_1}, \ldots, \vb{x}_{n_k}$ from a video using a sampler $\vb{\Theta}$. We encode the samples using a backbone $f$ followed by a two layer MLP, $g$ that creates a 128 dimensional embedding vector. We denote the embedding vectors as:
\begin{align}
    \vb{q} = g(f(\vb{x}_a)), \qquad
    \vb{z}^+ = g(f(\vb{x}_p)), ~ \text{and} \qquad
    \vb{z}_j^- = g(f(\vb{x}_{n_j})).
\end{align}
We maximize the agreement between (positive, anchor) and minimize between (anchor, negatives) pairs in the embedding space. Let $N$ be the set of cross video negatives generated from other videos in the dataset. To exploit the cross-video negatives, we calculate the normalized similarity measure between $\vb{q}$ and $\vb{z}$ for all $\vb{z} \in N$:
\begin{align}
    \alpha_{\vb{q},\vb{z}}= \frac{\exp\big(s(\vb{q}, \vb{z})/\tau\big)}{ \sum_{\vb{z}_c\in N}\exp\big(s(\vb{q}, \vb{z}_c)/\tau\big)},
\end{align}
where $s(\vb{a}, \vb{b}) = \vb{a}\cdot \vb{b}\big/(||\vb{a}||_2~ ||\vb{b}||_2)$ is the cosine similarity and $\tau$ is a temperature scaling parameter.
We then use the $\alpha_{\vb{q},\vb{z}}$ to rank the cross-video negatives according to their similarity with the anchor. Note that, with increasing similarity, the hardness of the negatives increase. We pick the top-$n$ hardest cross-video negatives, $N_m$, and compute $\vb{\hat{z}}^- = \sum_{\vb{z}\in N_m}{\alpha_{\vb{q},\vb{z}} \vb{z}}$. 
Finally, we minimize the loss,
\begin{align}
    \mathcal{L} = -\log \frac{\exp\big(s(\vb{q}, \vb{z}^+)/\tau\big)}{\exp\big(s(\vb{q}, \vb{z}^+)/\tau\big) + \sum_{j=1}^k\exp\big(s(\vb{q}, \vb{z}_j^-)/\tau\big) + \exp\big(s(\vb{q}, \vb{\hat{z}}^-)/\tau\big)}.
\label{eqn:loss_term}
\end{align}
Note that $\vb{z}_j^-$ is obtained intra-video, and $\vb{\hat{z}}^-$ is obtained from inter-video samples. Hence, $\mathcal{L}$ exploits both the intra-video and cross-video hard negatives. We chose $\tau\!=\!0.07$. Value of $n$ was $4$ and $2$ for GB videos and Butterfly, respectively.

\mypara{Video Sub-Sampling}
Most SOTA methods sample the anchor and positive pairs uniformly at random from the entire sequence of frames in a video. However, in the case of US videos, the view may change significantly if the samples are temporally distant. For example, in a transabdominal US video, one sample may show a GB with some parts of a liver while another may show only a liver and not a GB. Pairing such samples as positives would not work for learning a representation of the GB pathology. We recommend sampling the anchor and positives from a temporally close interval. We use a sampler, $\vb{\Theta}\!:\!\vb{V}\!\rightarrow\!\big(\vb{x}_a, \vb{x}_p, \{\vb{x}_{n_1}, \ldots, \vb{x}_{n_k}\}\big)$ to get the anchor ($\vb{x}_a$), positive ($\vb{x}_p$), and $k$ negative frames ($\vb{x}_{n_1}, \ldots, \vb{x}_{n_k}$) from a video $\vb{V}\!=\!\big\{ \vb{x}_j\big\}_{j=1}^M$. The indices of the anchor, positive, and negative frames are sampled as following: 
\begin{align*}
a \overset{\hphantom{\text{i.i.d.}}}{\sim} U\big([1,M]\big), 
\qquad
p \overset{\hphantom{\text{i.i.d.}}}{\sim} U\big([a-\delta, a+\delta] \setminus \{a\}\big), 
\\ 
n_1, \ldots, n_k \overset{\text{i.i.d.}}{\sim} U\big([1,M] \setminus [a-\Delta,a+\Delta]\big)
\label{eqn:neg_sample}
\end{align*} 
where $U(I)$ denotes sampling uniformly at random from interval $I$. We vary $\Delta$ between the $\Delta_h$ and $\Delta_l$ during the curriculum to adjust the hardness of the mined negatives. Also, $1 \le \delta \ll\ \!M$ and $ \Delta_{l} \le \Delta \le \Delta_{h} < M$. 

\mypara{Curriculum-based Negative Mining}
We use the negative samples in a hardness-sensitive order for effective learning. The model would initially learn to distinguish anchors from distant negatives and then gradually closer and thus harder negatives will be introduced. We start the training with only cross-video negatives and minimize the loss term, 
\begin{align}
    \mathcal{L}_{\text{cross}} = -\log \frac{\exp\big(s(\vb{q}, \vb{z}^+)/\tau\big)}{\exp\big(s(\vb{q}, \vb{z}^+)/\tau\big) + \exp\big(s(\vb{q}, \vb{\hat{z}}^-)/\tau\big)}.
\end{align}
We then gradually start using the loss in \cref{eqn:loss_term} to introduce intra-video negatives, which are more challenging to distinguish from the anchor than the cross-video negatives. We initially keep the $\Delta= \Delta_h =\lceil M/5\rceil$ used in \cref{eqn:neg_sample} for sampling the negatives and ensure the anchor and negatives are at least $\Delta$ frames apart temporally, and the hardness is comparatively lower. We gradually lower the $\Delta$ using a cosine annealing during the later phase of training to introduce harder negatives. We chose $\delta=3$, $k=3$, and  $\Delta_l=7$ in our experiments.

%
%
\section{Experiments and Results}

\begin{table}[t]
	\centering
	\setlength{\tabcolsep}{10pt}
	\caption{The fine-tuning performance of ResNet50 model in classifying malignant vs. non-malignant GBs from US images. We report accuracy, specificity, and sensitivity.}
    \resizebox{ \linewidth}{!}{%
	\begin{tabular}{lccc}
		\toprule
		\textbf{Method}	& \textbf{Acc.} & \textbf{Spec.} & \textbf{Sens.} \\
		\midrule
		Pretrained on \cite{imagenet} & 0.867 $\pm$ 0.070 & 0.926 $\pm$ 0.069 & 0.672 $\pm$ 0.147 \\
		\midrule
		SimCLR \cite{simclr} & 0.897 $\pm$ 0.040 & 0.912 $\pm$ 0.055 & 0.874 $\pm$ 0.067  \\
		SimSiam \cite{simsiam} & 0.900 $\pm$ 0.052 & 0.913 $\pm$ 0.059 & 0.861 $\pm$ 0.061 \\
		BYOL \cite{byol} & 0.844 $\pm$ 0.129 & 0.871 $\pm$ 0.144 & 0.739 $\pm$ 0.178 \\
		MoCo v2\cite{moco} & 0.886 $\pm$ 0.061 & 0.893 $\pm$ 0.078 & 0.871 $\pm$ 0.094 \\
		Cycle-Contrast \cite{cyclecontrast} & 0.861 $\pm$ 0.087 & 0.867 $\pm$ 0.098 & 0.844 $\pm$ 0.097 \\
		USCL \cite{uscl} & 0.901 $\pm$ 0.047 & 0.923 $\pm$ 0.041 & 0.831 $\pm$ 0.072 \\
		\midrule
		Ours & \textbf{0.921 $\pm$ 0.034} & \textbf{0.926 $\pm$ 0.043} & \textbf{0.900 $\pm$ 0.046} \\
		\bottomrule
	\end{tabular}
	}
	\label{tab:key_results}
\end{table}
\begin{table}
	\parbox{.5\linewidth}{
		\centering
		\setlength{\tabcolsep}{2pt}
		\caption{Comparison of finetuning performance of ResNet using the SOTA USCL, ImageNet pretraining, and our method on POCUS. We used the official finetuning script used by USCL. The pretraining was done on Butterfly dataset. The USCL official script reports the average accuracy over 5 runs. \textbf{C}, \textbf{P}, and \textbf{R} denote COVID-19, Pneumonia, and Regular respectively.}
		\begin{tabular}{lcccc}
		\toprule
		\multirow{2}{*}{\textbf{Method}} & \multicolumn{4}{c}{\textbf{Accuracy}} \\
		& \textbf{Overall} & \textbf{C} & \textbf{P} & \textbf{R} \\
		\midrule
		Pretrained \cite{imagenet} & 0.842 & 0.795 & 0.786 & 0.886 \\
		SimCLR & 0.864 & 0.832 & 0.894 & 0.871 \\
		MoCo v2 & 0.848 & 0.797 & 0.814 & 0.889 \\
		USCL & 0.907 & 0.861 & 0.903 & 0.935 \\
		Ours & \textbf{0.922} & \textbf{0.892} & \textbf{0.951} & 0.931 \\
		\bottomrule
		\end{tabular}
		\label{tab:pocus}
	}
	\hfill
	\parbox{.45\linewidth}{
		\centering
		\setlength{\tabcolsep}{3pt}
		\caption{We asked expert radiologists to classify GB malignancy for the test set of GBCU containing 80 non-malignant, and 42 malignant GB US images. Radiologists were not allowed access to any other patient data. The performance of the expert radiologists is comparable to that reported in the literature \cite{bo2019diagnostic,gupta2020evaluation}.}
		\begin{tabular}{lccc}
		\toprule
		\textbf{Method}	& \textbf{Acc.} & \textbf{Spec.} & \textbf{Sens.} \\
		\midrule
		Radiologist A & 0.816 & 0.873 & 0.707  \\
		Radiologist B & 0.784 & 0.811 & 0.732  \\
		\midrule
		USCL & 0.812 & 0.838 & 0.762 \\
		Pretrained \cite{imagenet} & 0.787 & 0.875 & 0.619 \\
		\midrule
		Ours & 0.877 & 0.900 & 0.833 \\
		\bottomrule
		\end{tabular}
		\label{tab:perf_human}
	}
\end{table}
\begin{table}[t]
	\centering
	\setlength{\tabcolsep}{10pt}
	\caption{Significance of joint mining of intra, and cross video negatives. While the individual mining techniques match SOTA performance in GB malignancy, pretraining with proposed joint mining  surpasses the current SOTA.}
	\resizebox{ \linewidth}{!}{%
	\begin{tabular}{ccccc}
		\toprule
		\multicolumn{2}{c}{\textbf{Type of negative used}}& \multirow{2}{*}{\textbf{Acc.}} & \multirow{2}{*}{\textbf{Spec.}} & \multirow{2}{*}{\textbf{Sens.}} \\
		Cross-video & Intra-video & & & \\
		\midrule
		 \checkmark & & 0.890 $\pm$ 0.062 & 0.897 $\pm$ 0.061 & 0.869 $\pm$ 0.108 \\
		 & \checkmark & 0.893 $\pm$ 0.057 & 0.904 $\pm$ 0.059 & 0.835 $\pm$ 0.109 \\
		 \checkmark & \checkmark & 0.921 $\pm$ 0.034 & 0.926 $\pm$ 0.043 & 0.900 $\pm$ 0.046 \\
		\bottomrule
	\end{tabular}
	}
	\label{tab:ablation_intra_loss}
\end{table}
\myfirstpara{Experimental Setup}
We use a machine with Intel Xeon Gold 5218@2.30GHz processor and 4 Nvidia Tesla V100 GPUs for our experiments. We pretrain a ResNet50 encoder using SGD with LR 0.003, weight decay $10^{-4}$, and momentum 0.9 for 60 epochs with batch size 32. We used a grid-search strategy to select the sampling hyper-parameters. We use a cosine annealing of the LR. The parameters of the anchor and positive encoders are updated using momentum contrast with momentum coefficient, $m\!=\!0.999$. Size of the queue for cross-video negative set is $|N|\!=\!96$ for GB videos and $|N|\!=\!66$ for Butterfly. We fine-tune for 30 epochs with batch size of 64 and an SGD optimizer with weight decay $5\!\cdot\!10^{-4}$. The remaining hyper-parameters are set similar to that of the pretraining phase.

\mypara{Comparison with SOTA}
We compare the ResNet50 \cite{resnet} backbone pretrained on our contrastive learning framework with ImageNet pretraining, SOTA UCL techniques SimCLR \cite{simclr}, SimSiam \cite{simsiam}, MoCo \cite{moco}, BYOL \cite{byol}, and SOTA image representation learning from video methods: Cycle-Contrast \cite{cyclecontrast} and USCL \cite{uscl}. USCL is specialized for pretraining on the US datasets. We note the performance of our pretraining framework for the GB malignancy classification task in \cref{tab:key_results}. Our method gives 92.1\% overall accuracy which is 2\% higher than SOTA and 90\% accuracy on malignant samples (sensitivity) which is 7\% higher than USCL. Our method also outperforms human experts significantly for detecting GB malignancy from US images (\cref{tab:perf_human}). We show the performance comparison on the POCUS dataset in \cref{tab:pocus} and observe that our method surpasses the SOTA USCL pretraining by 1.5\%. Fig. S1 in the supplementary material shows the Grad-CAM \cite{gradcam} visuals of the last conv layer to demonstrate that the attention regions of contrastive backbones are more precise and clinically relevant.

\mypara{Generality of our Method} 
We have shown our method's efficacy on two different tasks - (1) GB malignancy detection from abdominal US and (2) COVID-19 detection from lung US, which establishes the generality of our method on US modality. We also performed preliminary analysis on the performance of a ResNet50 classifier in detecting COVID-19 from a public CT dataset \cite{cov-finetune}. We pretrained the model on another CT dataset \cite{cov-pretrain}. The (accuracy, specificity, sensitivity) for our method was (0.80, 0.81, 0.80) as compared to (0.73, 0.72, 0.74) of ImageNet pretraining, and (0.78, 0.81, 0.76) of USCL. The results are indicative of the generality of our method across modalities. 

\begin{table}[t]
	\centering
	\setlength{\tabcolsep}{10pt}
	\caption{Effectiveness of our curriculum-based negative mining. The other alternative, curricula based trained models, lag significantly in GB malignancy classification.}
	\begin{tabular}{lccc}
		\toprule
		\textbf{Method}	& \textbf{Acc.} & \textbf{Spec.} & \textbf{Sens.} \\
		\midrule
		Proposed curriculum & 0.921 $\pm$ 0.034 & 0.926 $\pm$ 0.043 & 0.900 $\pm$ 0.046 \\
		Anti-curriculum & 0.887 $\pm$ 0.064 & 0.902 $\pm$ 0.056 & 0.836 $\pm$ 0.097 \\
		Control-curriculum & 0.897 $\pm$ 0.067 & 0.918 $\pm$ 0.062 & 0.810 $\pm$ 0.114 \\
		\bottomrule
	\end{tabular}
	\label{tab:ablation_curriculum}
\end{table}

\begin{figure}[t]
    \centering
    \includegraphics[width=\linewidth]{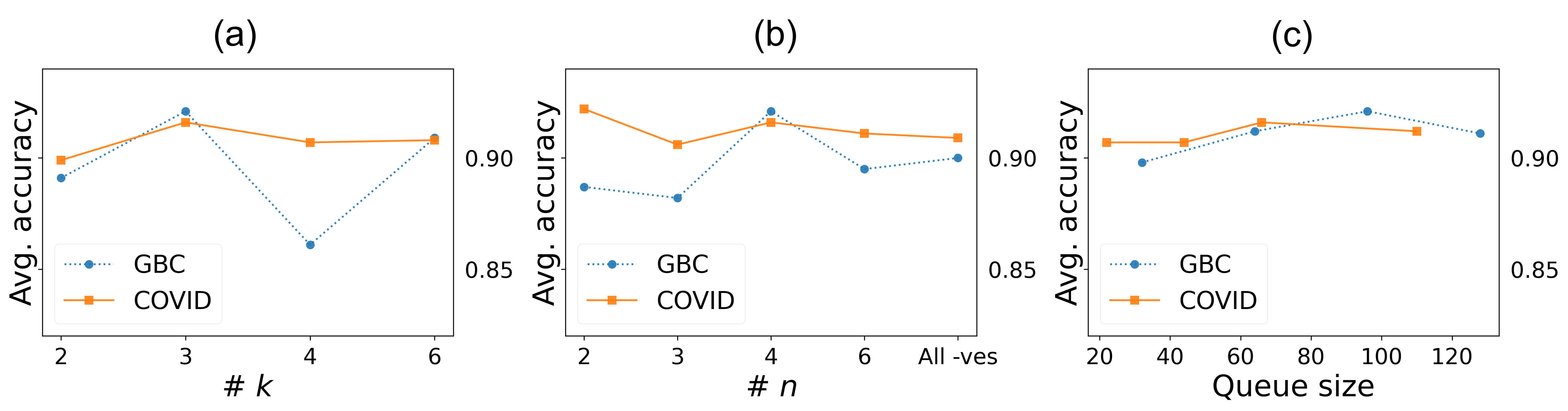}
    \caption{Sensitivity of pretraining hyper-parameters - (a) number of intra-video negatives ($k$), (b) number of top cross-video negatives ($n$), and (c) queue size ($|N|$) - on downstream accuracy. Mean cross-val accuracy is shown for GB Cancer and COVID.}
    \label{fig:sens-hyper}
\end{figure}

\mypara{Ablation Study}
\begin{enumerate*}[label=(\arabic*)]
	\item \textbf{Significance of Mining Intra-Video Hard Negatives:}
In \cref{tab:ablation_intra_loss} we observe that when the intra-video negatives are not used, the performance of malignancy detection of our method becomes comparable to that of Cycle-Contrast and USCL; both methods use only cross-video negatives. On the other hand, if only intra-video negatives are used, the model performance becomes similar to that of the SOTA image contrastive techniques. This shows the importance of mining both intra-video and cross-video negatives in achieving the performance boost.

\item \textbf{Effectiveness of Curriculum-based Negative Mining:}
We propose to use increasing order of hardness for negative mining during the training. To assert the effectiveness of such a curriculum, we compare the curriculum with two possible alternatives - (i) \emph{anti-curriculum} initially trains with harder negatives and progressively lowers the hardness of the negatives, and (ii) \emph{control-curriculum}  does not order the negatives according to their hardness. We initially train the model with temporally close intra-video negatives during the anti-curriculum and gradually start sampling temporally distant negatives. During the last few epochs, only the cross-video negatives are used. \cref{tab:ablation_curriculum} shows the performance comparison of the proposed hardness-sensitive curriculum over the two alternative curricula.

\item \textbf{Sensitivity of Hyper-parameters:}
\cref{fig:sens-hyper} shows the sensitivity of three important pretraining hyper-parameters on the accuracy of the downstream task for our method.
\end{enumerate*}

%
%
\section{Conclusion}
We introduce the first large-scale US video dataset for learning GB malignancy representation and propose an efficient UCL framework that exploits both intra-video and cross-video negatives through a hardness-aware curriculum. Our framework surpasses the human experts, imageNet-pretrained DNNs, and DNNs pretrained with SOTA contrastive learning methods specialized for US modality.

%
%
%
\bibliographystyle{splncs04}
\bibliography{Paper1638}

\section*{Supplementary Material}

\appendix
\beginsupplement
\section{Visualization of Feature Representation}
\begin{figure}[!h]
    \centering
    \includegraphics[width=\textwidth]{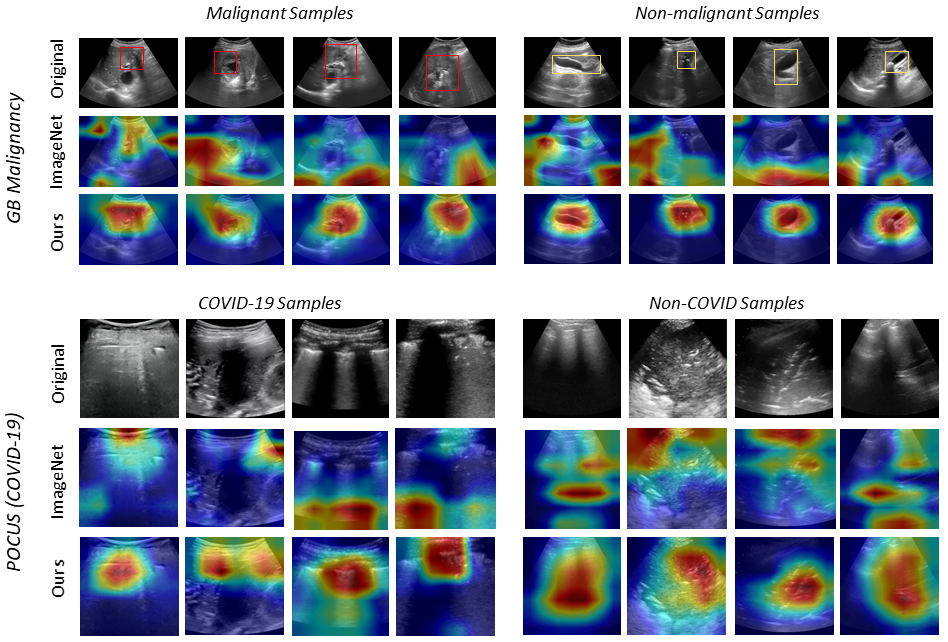}
    \caption{Grad-CAM visuals of the last conv layer in ImageNet pretrained model and the model pretrained using our CL method. The attention regions of contrastive backbones are more precise and clinically relevant.}
    \label{fig:cam_view}
\end{figure}

\cref{fig:cam_view} shows the Grad-CAM visuals using the features generated by the last convolutional layer. 
\section{Performance with Another Backbone}
We compare the fine-tuning performance of ResNet18 backbone by pretraining with our method is compared with ImageNet-pretrained ResNet18 and USCL-pretrained ResNet18. We show the results in \cref{tab:res18_gbc_results,tab:res18_pocus_results}. In the main text, we showed performance analysis with ResNet50 backbone. The superior performance of both backbones pretrained with our method indicates the generalizability of our framework to multiple backbones.
\begin{table}[!t]
    \parbox{\linewidth}{
    	\centering
    	\setlength{\tabcolsep}{10pt}
    	\caption{Finetuning performance of ResNet18 for classifying malignant vs. non-malignant GBs from USG images. Both ResNet50 (result in main text) and ResNet18 backbones show better accuracy and sensitivity of GB malignancy detection with contrastive pretraining as compared to the ImageNet pretraining.}
        \resizebox{ \linewidth}{!}{%
    	\begin{tabular}{lccc}
    		\toprule
    		\textbf{Method}	& \textbf{Acc.} & \textbf{Spec.} & \textbf{Sens.} \\
    		\midrule
    		Pretrained on ImageNet &  0.844 $\pm$ 0.053 & 0.856 $\pm$ 0.054 & 0.795 $\pm$ 0.097 \\
    		USCL & 0.896 $\pm$ 0.061 & 0.916 $\pm$ 0.066 & 0.833 $\pm$ 0.099 \\
    		\midrule
    		Ours & 0.907 $\pm$ 0.064 & 0.919 $\pm$ 0.072 & 0.862 $\pm$ 0.069 \\
    		\bottomrule
    	\end{tabular}
    	}
    	\label{tab:res18_gbc_results}
    }
    \vspace{2em}
    
    \parbox{\linewidth}{
        \centering
    	\setlength{\tabcolsep}{10pt}
    	\caption{Comparison of finetuning performance of ResNet18 using the SOTA USCL, ImageNet pretraining, and our method on POCUS. The pretraining was done on Butterfly dataset.}
    	\resizebox{ \linewidth}{!}{%
    	\begin{tabular}{lcccc}
        	\toprule
        	\multirow{2}{*}{\textbf{Method}} & \multicolumn{4}{c}{\textbf{Accuracy}} \\
        	& \textbf{Overall} & \textbf{COVID-19} & \textbf{Pneumonia} & \textbf{Regular} \\
        	\midrule
        	Pretrained on ImageNet & 0.874 & 0.797 & 0.874 & 0.919  \\
        	USCL & 0.914 & 0.916 & 0.940 & 0.906 \\
        	\midrule
        	Ours & 0.922 & 0.902 & 0.946 & 0.926   \\
        	\bottomrule
    	\end{tabular}
    	}
    	\label{tab:res18_pocus_results}
	}
\end{table}

\end{document}